# A new coupled computational method in conjunction with three-dimensional finite volume schemes for nonlinear coupled constitutive relations

Zhongzheng. Jiang, Weifang. Chen, Wenwen. Zhao

*College of Aeronautics and Astronautics, Zhejiang University, Hangzhou 310027, China*

Abstract:

It is widely believed that non-equilibrium effect plays a vital role in high-speed and rarefied gas flows and the accurate simulation of these flows is far beyond the capability of Navier-Stokes-Fourier (NSF) equations, which are based on small deviation from local thermodynamic equilibrium. To effectively solve this challenging work, B. C. Eu provided another way different from extended hydrodynamic equations (EHE) such as Grad's 13 moment equations and Burnett equations and proposed generalized hydrodynamic equations (GHE) which were consistent with the laws of irreversible thermodynamics. Based on Eu's equations, a new computational model, namely the nonlinear coupled constitutive relations (NCCR), was developed by R. S. Myong subsequently and had been applied successfully to one-dimensional shock wave structure and two-dimensional rarefied flows.

In this paper, based on Myong's work, an uncoupled computational method is extended for solving three-dimensional highly nonlinear coupled algebraic constitutive equations. However, the uncoupled solver will suffer from unstable deficits in some cases. Therefore, we propose a reliable coupled computational algorithm firstly for the complete solution of NCCR model. Three-dimensional upwind FVM scheme is adopted for the multidimensional conservation laws to gain insight into NCCR model's validity and potential in hypersonic rarefied gas flow cases.

Finally, the developed computational method is validated by several elaborately selected cases, such as hypersonic rarefied flows over a 2D cylinder covering slip and transitional regimes at 5.48 and 4 Mach numbers respectively as well as a slip complex flow over a 3D hollow cylinder-flare configuration. It is shown that the new coupled solver makes great improvement in stability and accuracy of computations compared with the uncoupled method. Moreover, NCCR model yields good solutions in better agreement with DSMC and experimental data in most studied cases than NSF equations. All results imply that the NCCR model with the coupled solver is capable of capturing the real physical flow properties away from equilibrium and demonstrates its great potential capability in further application.

*Ph.D. candidate, College of Aeronautics and Astronautics, Zhejiang University, Hangzhou, China 310027
†Professor, College of Aeronautics and Astronautics, Zhejiang University, Hangzhou, China 310027
‡Assistant Professor, College of Aeronautics and Astronautics, Zhejiang University, Hangzhou, China 310027, Corresponding author, E-mail: wwzhao@zju.edu.cn.

# 1    Introduction

Despite recent progress, how to predict aerodynamic characteristics of hypersonic vehicle accurately and effectively during reentry remains to be elucidated in the field of computational aerodynamics. High flight speed with rarefied environment results in multiscale non-equilibrium effects, which could be the main feature and cause of research difficulties in this field. Knudsen number, defined as the ratio between mean molecule free path and characteristic length, is the primary physical parameter that characterizes the rarefaction degree. Hypersonic reentry vehicles passing across different flight altitudes experience several flow regimes: continuum regime, slip regime, transition regime and free molecular flow regime, corresponded to $Kn \leq 0.01$ , $0.01 \leq Kn \leq 0.1$ , $0.1 \leq Kn \leq 10$ , $Kn \geq 10$ respectively. The well-known laws of Navier-Stokes-Fourier (NSF), which are based on the continuum assumption that the mean molecule free path is much smaller than the characteristic lengths of interest, are applicable only for flows at sufficiently small Knudsen number. As the flight altitude increases and the relevant length scale can be comparable to the mean free path, NSF equation will lose its superiority and hence more refined set of new theoretical tools of analysis beyond the classical theory of linear constitutive relations need to be developed.

In order to get the real physical solution of the challenging flow problems of which NSF equations fail in description, much work can be categorized into: 'particle simulation theory', 'gas kinetic theory' and 'fluid dynamic models'. In the former category, one of the most successful and well-known methods has to be the direct simulation Monte Carlo (DSMC) method based on tracking statistically representative particles to simulate the rarefied gas flow, which was first proposed by G. A. Bird in 1994[1]. Its great success in predicting the rarefied gas flow, particularly in free molecular flow regime and transition regime, has established its supreme status as an authority of standard for validating other methods in relative regimes. However, subjected to computational resource, its cost is rather prohibitive and unbearable in near continuum regime, even in the regimes away from continuum. To reduce the expensive computational consumption, some hybrid methods, such as NS/DSMC coupling method[2], have been put forward, but how to exchange information and when to switch between continuum and statistical methods still remains to be investigated.

Gas kinetic theory provides an alternative option to describe the non-equilibrium problems. Boltzmann equation, as the central equation in kinetic theory, describes the space-time evolution of a gas distribution function which contains the information of gas atoms at certain location with microscopic velocities, and describes the microscopic behavior of gases over a space scale of molecular mean free path. However, Boltzmann equation is a kind of integral-differential equation and its highly nonlinear particle collision term has greatly increased the complexity for directly solutions. Much effort has been put into the simplification of the collision term, such as linearization methods of Boltzmann equation[3], modeling equation methods[4], discrete velocity methods (DVM)[5] and discrete ordinate method (DOM)[6, 7],etc. In recent years, based on modeling equations, such as BGK, ES-BGK and Shakhov models, a unified gas-kinetic scheme(UGKS) for all flow regimes has been constructed with the discretization of particle velocity space by K. Xu, etc[8, 9]. UGKS overcomes the time step barrier for DSMC, direct Boltzmann solver and many other kinetic solvers and has gained its great success in homogeneous flow relaxation, shock structure calculations and flat plate flow problems[10].

However, in high speed flow with a wide spreading of particle velocity distribution, the requirement of a mesh in velocity space covering a large domain with high resolution makes the storage and computational load extremely unendurable, especially in three-dimensional hypersonic flows.

'Fluid dynamic models', including Burnett-type equations[11-13] that originate from the second-order Chapman-Enskog solution of Boltzmann equation, Grad's moment equations[14] derived by extended thermodynamics and Eu's moment equations[15] based on the generalized hydrodynamics, has been developed from Boltzmann equation for a dilute gas with emphasis on the computational efficiency and potential capability in engineering application. On the basis of the hyperbolic conservation laws of collision-free conserved variables, these methods deal with the non-equilibrium processes existing in flow regions of high local gradients of flow parameters. High-order constitutive relations in these methods, which express non-conserved variables in terms of conserved variables and their derivatives, are derived to describe the non-equilibrium effect. However, Grad's 13 moment equations and some of Burnett equations have been proved to be contradictory to Gibbs relation[16, 17] in some terms and their absence of the consistency with the second law of thermodynamics has manifested their defective capacity in capturing high-speed and low-density flow physics.

With a goal of providing a reliable high-order fluid dynamic model with stable computational capacity, Eu proposed a set of generalized hydrodynamic equations (GHE) from the viewpoint of generalized hydrodynamics formulated in the non-equilibrium ensemble method[18, 19]. He adopted extended irreversible thermodynamics as a tool to provide statistical mechanical foundations. The remarkable work of this theory is ingenious construction of non-equilibrium canonical distribution function as a bridge connecting entropy production with dissipative evolution of macroscopic non-conserved variables, which is strictly enforced to be consistent to the second law of thermodynamics by cumulant expansion for Boltzmann collision term. The first-order cumulant expansion takes a form of hyperbolic sine function, whose argument turns out to be in terms of a Rayleigh-Onsager dissipation function for non-conserved variables in near-equilibrium flow region. GHE has been successfully applied to calculate the shock profile for high Mach numbers[20] and to study sound wave absorption and dispersion in molecular gases[21].

However, when comes to multidimensional problems, GHE seems difficult to be put into the hyperbolic conservation system of partial differential equations, for which a variety of modern numerical methods are available, due to existence of the highly nonlinear coupled complicated terms for non-conserved variables. For this reason, Myong developed an efficient multidimensional computational model for GHE, which takes a form of nonlinear algebraic system and can be solved more easily coupled with the hyperbolic conservation laws by an iterative method[22]. This computational model has been validated by one-dimensional shock wave structure and two-dimensional flat plate flow problems for monatomic gases, and manifests its potential capacity in capturing the flow physics in high-speed and low-density flow regions[23]. Subsequently, by considering excess normal stress associated with the bulk viscosity of gases, this model was extended for a diatomic gas and has also been adopted successfully in two-dimensional hypersonic rarefied flow around a blunt body[24]. Since this model namely nonlinear coupled constitute relations (NCCR) has gained its success, more researches [25-27] have been carried out, including a discontinuous Galerkin method developed by H. Xiao and R.S.

Myong to ensure the high-order accuracy for NCCR model[28].

In present work, we aim to extend NCCR model to three-dimensional conservation laws problems and examine to what extent it can remain effective in complex hypersonic rarefied flow regions removed far from equilibrium. Myong's uncoupled computational method [23, 24] is extended firstly for the solution of three-dimensional highly nonlinear coupled algebraic constitutive equations. All cases are resolved by adopting finite volume schemes with various modern numerical techniques, such as LU-SGS implicit scheme for time advance, spatial discretization for inviscid flux by MUSCL interpolation and AUSMPW+ flux scheme, spatial discretization for viscous flux by central difference scheme. Our research shows that there are some limitations of computational stability for NCCR model computed by the uncoupled solver in some 3D flows. In order to overcome the computational instability, our main emphasis in this paper moves onto the direct solutions of the nonlinear coupled constitutive equations by a coupled solver.

The present paper is organized as follows. First, a brief deduction of the nonlinear coupled constitutive relations model from Eu's generalized hydrodynamic theory is summarized. Computational algorithms with upwind finite volume method and Maxwell-Smoluchowski boundary conditions are described in later section. The new coupled computational method for the solutions of the 3D nonlinear algebraic equations is also developed in this section. In Section 4, numerical results for hypersonic rarefied gas flows around a 2D cylinder and a 3D hollow cylinder flare configurations, are presented to demonstrate the capability of NCCR model in non-equilibrium flows and validate the developed solver. Section 5 summarizes the results of this investigation.

## 2    Deduction of three-dimensional NCCR model

A thermodynamically consistent extension of the Boltzmann equation to dilute rigid diatomic gases was done by Curtiss[29]. This irreversible kinetic equation, namely Boltzmann-Curtiss kinetic equation, which looks rather similar to the Boltzmann equation of monatomic gases, contains more terms about molecular rotation, under the assumption of no external field. The distribution function $f$ is not only a function of particle velocity $\mathbf{v}$, position $\mathbf{r}$ and time $t$, but also includes more information about internal rotational degree of freedom, such as the angular momentum $\mathbf{j}$ and the azimuthal angle $\psi$ associated with the orientation of the molecule. The Boltzmann-Curtiss kinetic equation is given by

$$(\frac{\partial}{\partial t}+\mathbf{v}\cdot\nabla+\mathbf{L}_r)f(\mathbf{v},\mathbf{r},\mathbf{j},\psi,t)=\mathrm{R}[f] , \tag{1}$$

where $\mathbf{L}_r$ is the internal Liouville operator expressed by

$$\mathbf{L}_r = \frac{\mathbf{j}}{I}\frac{\partial}{\partial \psi}+(\omega_B\times\mathbf{j})\cdot\frac{\partial}{\partial\mathbf{j}} , \tag{2}$$

among which $I$ stands for the moment of inertia and $\boldsymbol{\omega}_B$ denotes the vector of angular frequency of precession of the angular momentum. $\mathrm{R}[f]$ in Eq.(1) is the collision integral given by

$$\mathrm{R}[f]=\iiint\!\!\int\!\!\int d\mathbf{v}_1^*d\mathbf{v}_1 d\Omega_1 d\Omega^* d\Omega_1^* v_r^{'}\times\sigma\left(\mathbf{v}_r^{'},\mathbf{j}^*,\mathbf{j}_1^*|v_r,\mathbf{j},\mathbf{j}_1\right)\left(f^*f_1^* - ff_1\right) , \tag{3}$$

where asterisk stands for the post-collision value and $\mathbf{v}_r$ denotes the relative velocity. $d\Omega$ denotes the solid angle of scattering and $\sigma\left(\mathbf{v}_r^*, \mathbf{j}^*, \mathbf{j}_1^* \middle| \mathbf{v}_r, \mathbf{j}, \mathbf{j}_1\right)$ represents the collision cross section. In addition, there are two different sets of macroscopic variables: one set of conserved variables $\left(\rho, \mathbf{u}, E\right)$ and the other set of non-conserved variables $\left(\mathbf{\Pi}, \Delta, \mathbf{Q}\right)$, which can be defined by a unified statistical mechanical formula as

$$\Phi^{(k)} = \left\langle h^{(k)} f(\mathbf{v}, \mathbf{r}, \mathbf{j}, \psi, t) \right\rangle, \tag{4}$$

where the angular bracket means integration over the variables $\mathbf{v}$ and $\mathbf{j}$. $h^{(k)}$ is the molecular expressions for moments in two different groups. The leading elements of the set of conserved variables are defined as

$$\Phi^{(1)} = \rho, \ \Phi^{(2)} = \rho\mathbf{u}, \ \Phi^{(3)} = \rho e. \tag{5}$$

The molecular expressions corresponding to this set is

$$h^{(1)} = m, \ h^{(2)} = mu, \ h^{(3)} = 1/2\,mC^2 + H_{rot}. \tag{6}$$

The other set of non-conserved variables are given by the following expressions

$$\Phi^{(4)} = \mathbf{\Pi} = [\mathbf{P}]^{(2)}, \ \Phi^{(5)} = \Delta = \frac{1}{3}\mathrm{Tr}\,\mathbf{P} - p, \ \Phi^{(6)} = \mathbf{Q}. \tag{7}$$

And the set of molecular expressions is shown as

$$h^{(4)} = [m\mathbf{CC}]^{(2)}, \ h^{(5)} = \frac{1}{3}mC^2 - p/n, \ h^{(6)} = (\frac{1}{2}mC^2 + H_{rot} - \hat{h}m)\mathbf{C}. \tag{8}$$

Throughout this work, $\mathbf{\Pi}$, $\Delta$, $\mathbf{Q}$ represent the shear stress, the excess normal stress and the heat flux respectively, among which the excess normal stress, related to the bulk viscosity, is considered to take rotational non-equilibrium effect into account in special temperature regime where the rotational relaxation is much faster than hydrodynamic relaxations[24]. $n$ and $\hat{h}$ represent the number density of molecules and the enthalpy density per unit mass. The symbol $\left[\mathbf{A}\right]^{(2)}$ denotes the traceless symmetric part of the second-rank tensor $\mathbf{A}$, namely

$$\left[\mathbf{A}\right]^{(2)} = \frac{1}{2}\left(\mathbf{A} + \mathbf{A}^t\right) - \frac{1}{3}\mathbf{I}Tr\mathbf{A}. \tag{9}$$

And the stress tensor $\mathbf{P}$ consists of upper stresses through the relation as

$$\mathbf{P} = \left(p + \Delta\right)\mathbf{I} + \mathbf{\Pi}, \tag{10}$$

where $\mathbf{I}$ denotes the unit second-rank tensor and $p$ is the pressure.

It should be noted that the expressions as Eq.(6) are collision invariants and there is no dissipation term for the set of conserved variables. Thus, the kinetic equation (1) can be derived into a set of evolution equations of conserved variables according to the conservation laws, which are listed as following

Mass conservation law:

$$\frac{D\rho}{Dt} + \rho\nabla \cdot \mathbf{u} = 0, \tag{11}$$

Momentum conservation law:

$$\rho\frac{D\mathbf{u}}{Dt} + \nabla \cdot \mathbf{P} = 0, \tag{12}$$

Energy conservation law:

$$\rho \frac{De}{Dt} + \nabla \cdot \mathbf{Q} + \mathbf{P} : \nabla \mathbf{u} = 0 . \tag{13}$$

The unified macroscopic evolution equations for the set of non-conserved variables (7) can be derived by differentiating Eq.(4) with time and employing the kinetic equation (1). The details of the derivation process is presented in the literature[19] and the final equation is given by

$$\rho \frac{d\left(\Phi^{(k)} / \rho\right)}{dt} + \nabla \cdot \psi^{(k)} = \Lambda_k + Z_k , \tag{14}$$

where $\psi^{(k)}, \Lambda_k\ Z_k$, denote the flux of high-order moments, the dissipation term accounting for energy dissipation accompanying the irreversible process, the kinematic term from the hydrodynamic streaming effect, respectively. And they are defined by

$$\psi^{(k)} = \left\langle \mathbf{C} h^{(k)} f(\mathbf{v}, \mathbf{r}, \mathbf{j}, \psi, t) \right\rangle , \tag{15}$$

$$\Lambda_k = \left\langle h^{(k)} \, \mathrm{R}[f] \right\rangle , \tag{16}$$

$$Z_k = \left\langle f(\frac{\partial}{\partial t} + \mathbf{v} \cdot \nabla + \mathbf{L}_r) h^{(k)} \right\rangle . \tag{17}$$

In order to handle the complex dissipation term in the system through molecular collisions, Eu[19] constructed a non-equilibrium canonical distribution function which ensures the non-negativity property of entropy production as well as calortropy production in both physical and mathematical sense despite whatever approximations are made to the distribution function, to abridge the entropy production term and the dissipative evolution of macroscopic non-conserved variables. The non-equilibrium canonical $f$ is defined as

$$f = \exp\left[ -\frac{1}{k_B T} \left( \frac{1}{2} mC^2 + H_{rot} + \sum_{k=1}^{\infty} X_k h^{(k)} - \mu \right) \right] , \tag{18}$$

where $\mu$ is the normalization factor. $X_k$ are functions of macroscopic variables and occupy the status similar to the coefficients of Maxwell-Grad moment method. $T$, $k_B$, $m$ and $H_{rot}$ represent temperature, Boltzmann constant, molecular mass and rotational Hamiltonian of molecule respectively. By substituting Eq. (8) into Eq. (17) and inserting Eu's distribution function(18) into the dissipation term and employing the cumulant expansion method for Boltzmann collision term, a set of evolution equations for non-conserved variables for diatomic gases, namely Eu's generalized hydrodynamic equations, can be obtained as

$$\rho \frac{D(\mathbf{\Pi}/\rho)}{Dt} + \nabla \cdot \psi_4 = -2\left[\mathbf{\Pi} \cdot \nabla \mathbf{u}\right]^{(2)} - \frac{p}{\eta} \mathbf{\Pi} q(\kappa) - 2(p+\Delta)\left[\nabla \mathbf{u}\right]^{(2)} , \tag{19}$$

$$\rho \frac{D(\Delta/\rho)}{Dt} + \nabla \cdot \psi_5 = -2\gamma'(\Delta\mathbf{I}+\mathbf{\Pi}) : \nabla \mathbf{u} - \frac{2}{3}\gamma' p\nabla \cdot \mathbf{u} - \frac{2}{3}\gamma' \frac{p}{\eta_b} \Delta q(\kappa) , \tag{20}$$

$$\rho \frac{D(\mathbf{Q}/\rho)}{Dt} + \nabla \cdot \psi_6 = -\mathbf{\Pi} \cdot c_p \nabla T - \mathbf{Q} \cdot \nabla \mathbf{u} - \frac{pc_p T}{\lambda} \mathbf{Q} q(\kappa) - (p+\Delta)c_p T\nabla \ln T . \tag{21}$$

In these equations, the nonlinear dissipation factor $q(\kappa)$ takes a form of hyperbolic sine function of the Rayleigh-Onsager dissipation function, accounting for dissipative collision term (16), which is considered to be the cornerstone of Eu's modified moment method. Since the evolution equation is an open system of partial differential equations and there is no single closure theory founded on a firm theoretical justification, Eu[20] provided a closure different from Grad's closure to equations by

$$\psi_4 = \psi_5 = \psi_6 = 0 . \tag{22}$$

However, the existence of time term of non-conserved variables results in great computational difficulties in such a nonlinear system, implying an urgent need for a possible numerically solvable form by some approximations. Eu[19] considered that two sets of macroscopic variables in Eq. (5) and Eq. (7) vary on two different time scales: the non-conserved variables change faster than conserved variables and thus have already arrived at their steady state under such a longer time scale of evolution for conserved variables. On the other hand, the conserved variables will remain constant on the time scale of change in non-conserved variables. Therefore, a set of steady evolution equations for the non-conserved variables by simply omitting the substantial time derivative $d\left(\Phi^{(k)} / \rho\right)\big/ dt = 0$ , could be algebraically solved with the conserved variables held constant in them. This is the basic idea of adiabatic approximation. In the work of Myong[23, 24], the term $\mathbf{Q} \cdot \nabla \mathbf{u}$ is omitted from the constitutive relation(21) just for the sake of simplicity. Together with the application of Eu's closure and adiabatic approximation, a generalized hydrodynamic computational model for diatomic gases, namely NCCR model, can be summarized below as

$$-2\left[\mathbf{\Pi} \cdot \nabla \mathbf{u}\right]^{(2)} - \frac{p}{\eta} \mathbf{\Pi} q(\kappa) - 2(p+\Delta)\left[\nabla \mathbf{u}\right]^{(2)} = 0 , \tag{23}$$

$$-2\gamma'(\Delta \mathbf{I} + \mathbf{\Pi}) : \nabla \mathbf{u} - \frac{2}{3}\gamma' p \nabla \cdot \mathbf{u} - \frac{2}{3}\gamma' \frac{p}{\eta_b} \Delta q(\kappa) = 0 , \tag{24}$$

$$-\mathbf{\Pi} \cdot c_p \nabla T - \frac{p c_p T}{\lambda} \mathbf{Q} q(\kappa) - (p+\Delta) c_p T \nabla \ln T = 0 . \tag{25}$$

It should be mentioned that the effect of $\mathbf{Q} \cdot \nabla \mathbf{u}$ remains unknown and the investigation about this term still remains to be carried out. Currently, our research is still based on the NCCR model proposed by Myong as above.

# 3  Numerical methods
## 3.1  Governing equation and nonlinear coupled constitutive relations

In present work, the following dimensionless variables and parameters are used, corresponding to

$$x^* = \frac{x}{L_0}, y^* = \frac{y}{L_0}, z^* = \frac{z}{L_0}, u^* = \frac{u}{a_\infty}, v^* = \frac{v}{a_\infty}, w^* = \frac{w}{a_\infty}, p^* = \frac{p}{\rho_\infty a_\infty^2}, \rho^* = \frac{\rho}{\rho_\infty}, T^* = \frac{T}{T_\infty}, \eta^* = \frac{\eta}{\eta_\infty}, \eta_b^* = \frac{\eta_b}{\eta_\infty}, \lambda^* = \frac{\lambda}{\lambda_\infty}, E^* = \frac{E}{a_\infty^2},$$

$$h^* = \frac{h}{a_\infty^2}, t^* = \frac{t}{L_0 / a_\infty}, R^* = \frac{R}{a_\infty^2 / T_\infty}, c_p^* = \frac{c_p}{a_\infty^2 / T_\infty} = \frac{1}{\gamma - 1}, c_v^* = \frac{c_v}{a_\infty^2 / T}, \mathbf{\Pi}^* = \frac{\mathbf{\Pi}}{\eta_\infty a_\infty / L_0}, \Delta^* = \frac{\Delta}{\eta_\infty a_\infty / L_0}, \mathbf{Q}^* = \frac{\mathbf{Q}}{\lambda_\infty T_\infty / L_0}.$$

Here the asterisks denote dimensionless parameters and are omitted below for notational brevity. The dimensionless evolution equations for a diatomic gas can be summarized as

$$\frac{\partial \mathbf{U}}{\partial t} + \nabla \cdot \mathbf{F}_c + N_\delta \nabla \cdot \mathbf{F}_v = 0 , \tag{26}$$

$$\mathbf{U} = \begin{pmatrix} \rho \\ \rho\mathbf{u} \\ \rho E \end{pmatrix} \quad \mathbf{F}_c = \begin{pmatrix} \rho\mathbf{u} \\ \rho\mathbf{u}\mathbf{u} + p\mathbf{I} \\ (\rho E + p)\mathbf{u} \end{pmatrix} \quad \mathbf{F}_v = \begin{pmatrix} 0 \\ \mathbf{\Pi} + \Delta\mathbf{I} \\ (\mathbf{\Pi} + \Delta\mathbf{I}) \cdot \mathbf{u} + \varepsilon\mathbf{Q} \end{pmatrix},$$

and

$$\hat{\mathbf{\Pi}} q\left(c\hat{R}\right) = (1 + \hat{\Delta})\hat{\mathbf{\Pi}}_0 + \left[\hat{\mathbf{\Pi}} \cdot \nabla\hat{\mathbf{u}}\right]^{(2)}, \tag{27}$$

$$\hat{\Delta} q\left(c\hat{R}\right) = \hat{\Delta}_0 + \frac{3}{2} f_b (\hat{\Delta}\mathbf{I} + \hat{\mathbf{\Pi}}) : \nabla\hat{\mathbf{u}}, \tag{28}$$

$$\hat{\mathbf{Q}} q\left(c\hat{R}\right) = (1 + \hat{\Delta})\hat{\mathbf{Q}}_0 + \hat{\mathbf{\Pi}} \cdot \hat{\mathbf{Q}}_0, \tag{29}$$

where $\hat{\mathbf{\Pi}}_0$, $\hat{\Delta}_0$ and $\hat{\mathbf{Q}}_0$ represent the linear Newtonian law of shear stress, linear excess normal stress and the Fourier law of heat conduction, respectively,

$$\mathbf{\Pi}_0 = -2\eta\left[\nabla\mathbf{u}\right]^{(2)}, \quad \Delta_0 = -\eta_b\nabla\cdot\mathbf{u}, \quad \mathbf{Q}_0 = -\lambda\nabla T, \tag{30}$$

where

$$\frac{\eta_\infty}{\rho_\infty a_\infty L_0} = N_\delta = \frac{\mathrm{Ma}}{\mathrm{Re}}, \quad \hat{\mathbf{\Pi}} = \frac{N_\delta}{p}\mathbf{\Pi}, \quad \hat{\Delta} = \frac{N_\delta}{p}\Delta, \quad \hat{\mathbf{Q}} = \frac{N_\delta}{p}\frac{\mathbf{Q}}{\sqrt{T/(2\varepsilon)}}$$

$$\hat{R} = \left[\hat{\mathbf{\Pi}} : \hat{\mathbf{\Pi}} + \frac{2\gamma'}{f_b}\hat{\Delta}^2 + \hat{\mathbf{Q}}\cdot\hat{\mathbf{Q}}\right]^{1/2}, \nabla\hat{\mathbf{u}} = -2\eta\frac{N_\delta}{p}\nabla\mathbf{u}, \varepsilon = \frac{1}{\mathrm{Pr}(\gamma-1)}$$

The viscosity $\eta$ is modeled with the inverse power laws as

$$\eta = \frac{5m\left(RT/\pi\right)^{\frac{1}{2}}\left(2mRT/K\right)^{\frac{2}{\nu-1}}}{8A_2\left(\nu\right)\Gamma\left[4 - \frac{2}{\nu-1}\right]}. \tag{31}$$

And the formula (31) of the inverse power laws can actually be defined as

$$\eta = \eta_{ref}\left(\frac{T}{T_{ref}}\right)^s, \tag{32}$$

where $s$ is given by $s = 1/2 + 2/(\nu-1)$. The parameter $\nu$ is the exponent of the inverse power laws. The nonlinear dissipative factor is defined by $q\left(c\hat{R}\right) = \sinh\left(c\hat{R}\right)/c\hat{R}$ and the constant $c$ is given by

$$c = \left(mk_B T_r\right)^{\frac{1}{4}}\frac{1}{2d_r\eta_r^{1/2}}.$$

Based on the inverse power laws in Eq. (31) above, the constant $c$ reduces to a function of the exponent of inverse power laws, given by

$$c = \left[\frac{2\sqrt{\pi}}{5}A_2\left(\nu\right)\Gamma(4 - \frac{2}{\nu-1})\right]^{\frac{1}{2}}. \tag{33}$$

The values of function $A_2\left(\nu\right)$ can be obtained in the literature[30] and the gamma function $\Gamma$ is computed from a table list in the literature[31]. For a perfect gas, the following dimensionless relations hold

$$p = \rho T / \gamma, c_p = 1 / (\gamma - 1), d = T^{1/(1-\nu)}$$

$$\eta = T^s, \eta_b = f_b \eta = f_b T^s, \lambda = \eta = T^s \quad,$$

$$T^{1/4} / (d \sqrt{\eta}) = 1, \gamma' = (5 - 3\gamma) / 2$$

where $f_b$ denotes the ratio of the bulk viscosity to the shear viscosity and its value for nitrogen gas sets 0.8 according to the literature[21].

## 3.2   Spatial discretization and time integration

The integral form of the governing equations (26) based on a finite volume $\Omega$ can be expressed as

$$\iiint \frac{\partial \mathbf{U}}{\partial \tau} d\Omega + \oiint_{\partial \Omega} \left( \mathbf{F}_c + N_\delta \mathbf{F}_v \right) \cdot \mathbf{n} ds = \iiint \frac{\partial \mathbf{U}}{\partial \tau} d\Omega + \oiint_{\partial \Omega} \left( \mathbf{F} n_x + \mathbf{G} n_y + \mathbf{H} n_z \right) ds = 0, \tag{34}$$

where

$$\mathbf{F}_c = F_c \mathbf{i} + G_c \mathbf{j} + H_c \mathbf{k}, \ \mathbf{F}_v = N_\delta F_v \mathbf{i} + N_\delta G_v \mathbf{j} + N_\delta H_v \mathbf{k}$$

$$\mathbf{F} = \left( F_c + N_\delta F_v \right) \mathbf{i}, \ \mathbf{G} = \left( G_c + N_\delta G_v \right) \mathbf{j}, \ \mathbf{H} = \left( H_c + N_\delta H_v \right) \mathbf{k} \quad.$$

And the semi-discrete scheme for the governing equations is introduced by

$$\frac{\partial \mathbf{U}_{I,J,K}}{\partial \tau} + \mathbf{F}_{I+1/2} - \mathbf{F}_{I-1/2} + \mathbf{G}_{J+1/2} - \mathbf{G}_{J-1/2} + \mathbf{H}_{K+1/2} - \mathbf{H}_{K-1/2} = 0. \tag{35}$$

In order to numerically solve the discrete equation above, the modern CFD upwind schemes based on the hyperbolic conservation laws can be adopted. Since AUSMPW+ flux scheme has great capacity in capturing shock wave and contact discontinuity precisely, it is employed to compute the inviscid flux in present work.

Note that accurate description of viscosity shows its decisive effects in high-speed and low-density flows based on fluid dynamic models. Therefore, a compressible limiter with less dissipative effects is preferable. In order to ensure higher order spatial accuracy, the MUSCL (Monotone Upstream-Centered Scheme for Conservation Laws) interpolation in conjunction with Van Albada limiter is employed as

$$q_{i+1/2}^L = q_i + \frac{1}{4} \left[ (1 - \kappa) \overline{\Delta}_- + (1 + \kappa) \overline{\Delta}_+ \right]_i, \tag{36}$$

$$q_{i+1/2}^R = q_{i+1} - \frac{1}{4} \left[ (1 - \kappa) \overline{\Delta}_+ + (1 + \kappa) \overline{\Delta}_- \right]_{i+1}, \tag{37}$$

where

$$\overline{\Delta}_- = \overline{\Delta}_+ = Van \ albada \left( \Delta_-, \Delta_+ \right)$$

$$\left( \Delta_- \right)_i = q_i - q_{i-1}, \ \left( \Delta_+ \right)_i = q_{i+1} - q_i \quad.$$

$$Van \ albada \left( x, y \right) = \frac{x \left( y^2 + \varepsilon \right) + y \left( x^2 + \varepsilon \right)}{x^2 + y^2 + 2\varepsilon}$$

And $\varepsilon$ is a small quantity in case of denominator's division by zero. Viscous term is discretized using two-order central difference scheme and the final semi-discrete iteration is advanced by LU-SGS (Lower-Upper Symmetric Gauss-Seidel) implicit scheme. It turns out that the following time step is available and stable for NCCR model, which is expressed as

$$\Delta t = \frac{CFL \cdot \Omega}{\lambda_{c,\xi} + \lambda_{c,\eta} + \lambda_{c,\zeta} + \dfrac{2}{Re_{ref}}\left(\lambda_{v,\xi} + \lambda_{v,\eta} + \lambda_{v,\zeta}\right)} \ . \tag{38}$$

The spectral radii for inviscid and viscous terms is given by

$$\lambda_{c,\xi} + \lambda_{c,\eta} + \lambda_{c,\zeta} = \frac{1}{J}\Big[|U| + |V| + |W| + c\left(|\nabla\xi| + |\nabla\eta| + |\nabla\zeta|\right)\Big],$$

$$\lambda_{v,\xi} + \lambda_{v,\eta} + \lambda_{v,\zeta} = \frac{2\mu}{J\rho}\left(|\nabla\xi|^2 + |\nabla\eta|^2 + |\nabla\zeta|^2\right)\max\left(\frac{4}{3}, \frac{\gamma}{\mathrm{Pr}}\right),$$

where $\Omega$ denotes the cell volume and $c$ represents the speed of sound.

### 3.3 Boundary conditions

Proper slip-wall boundary conditions are vital for rarefied flow computation. In present research, classical Maxwell-Smoluchowski boundary conditions, are employed for the NCCR model, which can be expressed as

$$u_{y=0} - u_w = C_m\lambda\frac{\partial u_0}{\partial y} + C_s\frac{\mu}{\rho T}\frac{\partial T}{\partial x},$$

$$T_{y=0} - T_w = C_t\lambda\frac{\partial T}{\partial y} \ , \tag{39}$$

where $C_m$, $C_s$, $C_t$ and $\lambda$ denotes velocity slip coefficient, hot creep coefficient, temperature jump coefficient and molecular mean free path respectively.

### 3.4 Iterative solutions of nonlinear coupled constitutive relations

#### 3.4.1 Uncoupled solver

Although NCCR model (27)-(29) has been simplified greatly from GHE, its numerical solution is still hard to be obtained due to its nonlinearity and coupled relations of the high-order non-conserved variables. Therefore, in order to solve the discretized equation (35) of the conservation laws, a nonlinear algebraic system of NCCR model was proposed by Myong[23, 24], which can be solved by an iterative method. The iterative solutions of NCCR model do not need to solve a hyperbolic system with more higher-order variables like aforementioned moment methods such as Grad's moment equations, but only require an additional process to calculate the stress and heat flux from the nonlinear algebraic system and then adopt them back into the conserved equation (26) to solve the conserved variables, which shares a similar feature with the Navier-Stokes-Fourier method solving five components of conserved moments. Based on Myong's two-dimensional algorithm, three-dimensional problem can be simplified approximately into three one-dimensional problems in x, y, z directions, where the stress and heat flux components $\left(\Pi_{xx}, \Pi_{xy}, \Pi_{xz}, \Delta, Q_x\right)$ on a surface in three-dimensional finite volume induced by thermodynamic forces $\left(u_x, v_x, w_x, T_x\right)$ can be approximated as the sum of two uncoupled solvers: one on $\left(u_x, 0, 0, T_x\right)$ describing the compression and expansion flows, and another on $\left(0, v_x, 0, 0\right)$ and $\left(0, 0, w_x, 0\right)$ describing shear flows. In order to give a brief uncoupled solution process for 3D NCCR model, we take a unified notation here and a rotation index is introduced firstly in the table below.

Table 1 description of the unified notation and rotation index

| $i$ | $x_i$ | $u_i$ | $j$ | $x_j$ | $u_j$ | $k$ | $x_k$ | $u_k$ |
|-----|-------|-------|-----|-------|-------|-----|-------|-------|
| 1 | $x$ | $u$ | 2 | $y$ | $v$ | 3 | $z$ | $w$ |
| 2 | $y$ | $v$ | 3 | $z$ | $w$ | 1 | $x$ | $u$ |
| 3 | $z$ | $w$ | 1 | $x$ | $u$ | 2 | $y$ | $v$ |

The first uncoupled solver in the normal direction $i$ of $x_i$ plane is given by

$$q\left(c\hat{R}\right)\hat{\Pi}_{x_i x_i}^{u_{i,i}} = \left(1 + \hat{\Delta}^{u_{i,i}} + \hat{\Pi}_{x_i x_i}^{u_{i,i}}\right)\hat{\Pi}_{x_i x_i 0}^{u_{i,i}}$$
$$q\left(c\hat{R}\right)\hat{\Delta}^{u_{i,i}} = \left[1 + 3\left(\hat{\Pi}_{x_i x_i}^{u_{i,i}} + \hat{\Delta}^{u_{i,i}}\right)\right]\hat{\Delta}_0^{u_{i,i}} , \tag{40}$$
$$q\left(c\hat{R}\right)\hat{Q}_{x_i} = \left(1 + \hat{\Delta}^{u_{i,i}} + \hat{\Pi}_{x_i x_i}^{u_{i,i}}\right)\hat{Q}_{x_i 0}$$

where

$$\hat{R}^2 = \frac{3}{2}\left(\hat{\Pi}_{x_i x_i}^{u_{i,i}}\right)^2 + \frac{2\gamma'}{f_b}\left(\hat{\Delta}^{u_{i,i}}\right)^2 + \hat{Q}_{x_i}^2 . \tag{41}$$

The second uncoupled solver in the two shear directions $j$ and $k$ of $x_i$ plane is expressed by

$$q\left(c\hat{R}\right)\hat{\Pi}_{x_i x_j}^{u_{j,i}} = -\frac{2}{3}\hat{\Pi}_{x_i x_i}^{u_{j,i}}\hat{\Pi}_{x_i x_j 0}^{u_{j,i}} \qquad q\left(c\hat{R}\right)\hat{\Pi}_{x_i x_k}^{u_{k,i}} = -\frac{2}{3}\hat{\Pi}_{x_i x_k}^{u_{k,i}}\hat{\Pi}_{x_i x_k 0}^{u_{k,i}}$$
$$q\left(c\hat{R}\right)\hat{\Pi}_{x_i x_j}^{u_{j,i}} = \left(1 + \hat{\Delta}^{u_{j,i}} + \hat{\Pi}_{x_i x_i}^{u_{j,i}}\right)\hat{\Pi}_{x_i x_j 0}^{u_{j,i}} \qquad q\left(c\hat{R}\right)\hat{\Pi}_{x_i x_k}^{u_{k,i}} = \left(1 + \hat{\Delta}^{u_{k,i}} + \hat{\Pi}_{x_i x_i}^{u_{k,i}}\right)\hat{\Pi}_{x_i x_k 0}^{u_{k,i}} , \tag{42}$$
$$q\left(c\hat{R}\right)\hat{\Delta}^{u_{j,i}} = 3f_b\hat{\Pi}_{x_i x_j}^{u_{j,i}}\hat{\Pi}_{x_i x_j 0}^{u_{j,i}} \qquad q\left(c\hat{R}\right)\hat{\Delta}^{u_{k,i}} = 3f_b\hat{\Pi}_{x_i x_k}^{u_{k,i}}\hat{\Pi}_{x_i x_k 0}^{u_{k,i}}$$

where

$$\hat{R}^2 = 6\left(\hat{\Pi}_{x_i x_i}^{u_{j,i}}\right)^2 + 2\left(\hat{\Pi}_{x_i x_i}^{u_{j,i}}\right)^2 + \frac{2\gamma'}{f_b}\left(\hat{\Delta}^{u_{j,i}}\right)^2 \qquad \hat{R}^2 = 6\left(\hat{\Pi}_{x_i x_i}^{u_{k,i}}\right)^2 + 2\left(\hat{\Pi}_{x_i x_i}^{u_{k,i}}\right)^2 + \frac{2\gamma'}{f_b}\left(\hat{\Delta}^{u_{k,i}}\right)^2 . \tag{43}$$

Here some symbols like $\hat{\Pi}_{x_i x_i}^{u_{i,i}}$, $\hat{\Pi}_{x_i x_i}^{u_{j,i}}$ need to be illuminated clearly. Take $\hat{\Pi}_{x_i x_i}^{u_{i,i}}$ for an example. The superscript $u_{i,i}$ denotes $\partial u_i / \partial x_i$ and expresses the partial derivative of velocity $u$ with respect to x-coordinate particularly in the case of $i = 1$ according to Table 1. $\hat{\Pi}_{x_i x_i}^{u_{j,i}}$ represents the shear stress in $x_i$ direction of $x_i$ plane induced by the gradient of velocity $\partial u_i / \partial x_i$.

The formulas (40)-(43) compose the nonlinear algebraic system of NCCR model (27)-(29). An iterative method, which is developed to solve the constitutive relations, can be summarized as follows.

The first uncoupled solver on $\left(u_{i,i}, 0, 0, T_{,i}\right)$:

The initial values of shear stress, excess normal stress and heat flux in NCCR model are calculated below by the linear values $\Pi_0$, $\Delta_0$ and $Q_0$ from NSF model as

$$\hat{R}_0^2 = \frac{3}{2}\left(\hat{\Pi}_{x_i x_i 0}^{u_{j,i}}\right)^2 + \frac{2\gamma'}{f_b}\left(\hat{\Delta}_0^{u_{j,i}}\right)^2 + \hat{Q}_{x_i 0}^2$$

$$\hat{\Pi}_{x_i x_i 1}^{u_{j,i}} = \frac{\sinh^{-1}\left(c\hat{R}_0\right)}{c\hat{R}_0}\hat{\Pi}_{x_i x_i 0}^{u_{j,i}}$$

$$\hat{\Delta}_1^{u_{j,i}} = \frac{\sinh^{-1}\left(c\hat{R}_0\right)}{c\hat{R}_0}\hat{\Delta}_0^{u_{j,i}} \qquad\qquad (44)$$

$$\hat{Q}_{x_i 1} = \frac{\sinh^{-1}\left(c\hat{R}_0\right)}{c\hat{R}_0}\hat{Q}_{x_i 0}$$

For positive $\hat{\Pi}_{x_i x_i 0}^{u_{j,i}}$ and $\hat{Q}_{x_i 0}$, the solution process is given by

$$\hat{R}_n^2 = \frac{3}{2}\left(\hat{\Pi}_{x_i x_i}^{u_{j,i}}\right)_n^2 + \frac{2\gamma'}{f_b}\left(\hat{\Delta}^{u_{j,i}}\right)_n^2 + \hat{Q}_{x_i n}^2$$

$$\hat{R}_{n+1} = \frac{1}{c}\sinh^{-1}\left[c\sqrt{Y_n}\right]$$

$$\left(\hat{\Pi}_{x_i x_i}^{u_{j,i}}\right)_{n+1} = \left[1 + \left(\hat{\Delta}^{u_{j,i}}\right)_n + \left(\hat{\Pi}_{x_i x_i}^{u_{j,i}}\right)_n\right]\hat{\Pi}_{x_i x_i 0}^{u_{j,i}}\frac{\hat{R}_{n+1}}{\sqrt{Y_n}} \qquad\qquad (45)$$

$$\left(\hat{Q}_{x_i}\right)_{n+1} = \frac{\hat{Q}_{x_i 0}}{\hat{\Pi}_{x_i x_i 0}^{u_{j,i}}}\left(\hat{\Pi}_{x_i x_i}^{u_{j,i}}\right)_{n+1}$$

$$\left(\hat{\Delta}^{u_{j,i}}\right)_{n+1} = \frac{1}{8}\left[\left(9f_b - 4\right)\left(\hat{\Pi}_{x_i x_i}^{u_{j,i}}\right)_{n+1} - 4 + \sqrt{D_{n+1}}\right]$$

where

$$D_{n+1} = \left(81f_b^2 + 72f_b + 16\right)\left(\hat{\Pi}_{x_i x_i}^{u_{j,i}}\right)_{n+1}^2 + \left(32 - 24f_b\right)\left(\hat{\Pi}_{x_i x_i}^{u_{j,i}}\right)_{n+1} + 16$$

$$Y_n = \left(1 + \left(\hat{\Delta}^{u_{j,i}}\right)_n + \left(\hat{\Pi}_{x_i x_i}^{u_{j,i}}\right)_n\right)^2\hat{R}_0^2 + 4\left[\left(\hat{\Pi}_{x_i x_i}^{u_{j,i}}\right)_n + \left(\hat{\Delta}^{u_{j,i}}\right)_n\right]\left[1 + 2\left[\left(\hat{\Pi}_{x_i x_i}^{u_{j,i}}\right)_n + \left(\hat{\Delta}^{u_{j,i}}\right)_n\right]\right]\frac{2\gamma'}{f_b}\left(\hat{\Delta}_0^{u_{j,i}}\right)^2 \qquad (46)$$

And for negative $\hat{\Pi}_{x_i x_i 0}^{u_{j,i}}$ and $\hat{Q}_{x_i 0}$, the solution process is given by

$$\hat{R}_n^2 = \frac{3}{2}\left(\hat{\Pi}_{x_i x_i}^{u_{j,i}}\right)_n^2 + \frac{2\gamma'}{f_b}\left(\hat{\Delta}^{u_{j,i}}\right)_n^2 + \hat{Q}_{x_i n}^2$$

$$\left(\hat{\Pi}_{x_i x_i}^{u_{j,i}}\right)_{n+1} = \frac{\left[1 + \left(\hat{\Delta}^{u_{j,i}}\right)_n\right]\hat{\Pi}_{x_i x_i 0}^{u_{j,i}}}{q\left(c\hat{R}_n\right) - \hat{\Pi}_{x_i x_i 0}^{u_{j,i}}}$$

$$\left(\hat{Q}_{x_i}\right)_{n+1} = \frac{\hat{Q}_{x_i 0}}{\hat{\Pi}_{x_i x_i 0}^{u_{j,i}}}\left(\hat{\Pi}_{x_i x_i}^{u_{j,i}}\right)_{n+1} \qquad\qquad (47)$$

$$\left(\hat{\Delta}^{u_{j,i}}\right)_{n+1} = \frac{1}{8}\left[\left(9f_b - 4\right)\left(\hat{\Pi}_{x_i x_i}^{u_{j,i}}\right)_{n+1} - 4 + \sqrt{D_{n+1}}\right]$$

Here $D_{n+1}$ can also be computed by formula (46).

The second uncoupled solvers on $\left(0, u_{j,i}, 0, 0\right)$ and $\left(0, 0, u_{k,i}, 0\right)$ are similar and thus only the process of the second solver on $\left(0, u_{j,i}, 0, 0\right)$ is presented below. Note that the stress $\Pi_{x_i x_i 0}$ deduced by thermodynamic forces $u_{j,i}$ does not exist in NSF linear constitutive relations. Therefore, the initial value $\hat{\Pi}_{x_i x_i 1}^{u_{j,i}}$ is equal to zero. From Eq. (42) a stress constraint can be obtained as

$$\hat{\Pi}_{x_i x_j}^{u_{j,i}} = sign(\hat{\Pi}_{x_i x_j 0}^{u_{j,i}})\left[-\frac{3}{2}\hat{\Pi}_{x_i x_i}^{u_{j,i}}\left(\left(1-\frac{9}{2}f_b\right)\hat{\Pi}_{x_i x_i}^{u_{j,i}}+1\right)\right]^{1/2},\tag{48}$$

whose mathematical characteristic will change from ellipse to hyperbola at critical point $2/9$ of the parameter $f_b$. Therefore, two situations are necessary to be distinguished in the following iterative method. For $f_b > 2/9$, the second uncoupled solver is given by

$$\hat{R}_n = \left\{3\left(\hat{\Pi}_{x_i x_i}^{u_{j,i}}\right)_n\left[\left(1+\frac{9}{2}(3\gamma'+1)f_b\right)\left(\hat{\Pi}_{x_i x_i}^{u_{j,i}}\right)_n-1\right]\right\}^{1/2}$$
$$\hat{R}_{n+1} = \frac{1}{c}\sinh^{-1}\left(c Y_n\right) ,\tag{49}$$
$$\left(\hat{\Pi}_{x_i x_i}^{u_{j,i}}\right)_{n+1} = \frac{3-\sqrt{D_{n+1}}}{3\left(2+9(3\gamma'+1)f_b\right)}$$

where

$$D_{n+1} = 9+12\left(1+\frac{9}{2}(3\gamma'+1)f_b\right)\hat{R}_{n+1}^2$$
$$Y_n = \left\{2\left[1+\left(1-\frac{9}{2}f_b\right)\left(\hat{\Pi}_{x_i x_i}^{u_{j,i}}\right)_n\right]\left[1-\left(1+\frac{9}{2}(3\gamma'+1)f_b\right)\left(\hat{\Pi}_{x_i x_i}^{u_{j,i}}\right)_n\right]\right\}^{1/2}\hat{\Pi}_{x_i x_j 0}^{u_{j,i}}.$$

For $f_b < 2/9$, the second uncoupled solver is given by

$$\left(\hat{\Pi}_{x_i x_i}^{u_{j,i}}\right)_{n+1} = \frac{-2\left(\hat{\Pi}_{x_i x_j 0}^{u_{j,i}}\right)^2}{3q^2\left(c\hat{R}_n\right)+\left(2-9f_b\right)\left(\hat{\Pi}_{x_i x_j 0}^{u_{j,i}}\right)^2}$$
$$\hat{R}_n = \left\{3\left(\hat{\Pi}_{x_i x_i}^{u_{j,i}}\right)_n\left[\left(1+\frac{9}{2}(3\gamma'+1)f_b\right)\left(\hat{\Pi}_{x_i x_i}^{u_{j,i}}\right)_n-1\right]\right\}^{1/2}.\tag{50}$$

When a converged value of $\hat{\Pi}_{x_i x_i}^{u_{j,i}}$ is obtained in the two processes above, it can be substituted into the following formulas to get new values of $\hat{\Delta}^{u_{j,i}}$ and $\hat{\Pi}_{x_i x_j}^{u_{j,i}}$ as

$$\left(\hat{\Delta}^{u_{j,i}}\right)_{n+1} = -\frac{9}{2}f_b\left(\hat{\Pi}_{x_i x_i}^{u_{j,i}}\right)_{n+1}$$
$$\left(\hat{\Pi}_{x_i x_j}^{u_{j,i}}\right)_{n+1} = sign(\hat{\Pi}_{x_i x_j 0}^{u_{j,i}})\left[-\frac{3}{2}\left(\hat{\Pi}_{x_i x_i}^{u_{j,i}}\right)_{n+1}\left(\left(1-\frac{9}{2}f_b\right)\left(\hat{\Pi}_{x_i x_i}^{u_{j,i}}\right)_{n+1}+1\right)\right]^{1/2}.\tag{51}$$

The iterative solutions of NCCR model are considered to be converged when $\left|\hat{R}_{n+1}-\hat{R}_n\right| \leq 10^{-5}$. After converged, all non-conserved variables are counted up together below at this time step as

$$\Pi_{xx} = \frac{p}{N_\delta}\left(\hat{\Pi}_{xx}^{u_x}+\hat{\Pi}_{xx}^{v_x}+\hat{\Pi}_{xx}^{w_x}\right) \quad \Pi_{xy} = \Pi_{yx} = \frac{p}{N_\delta}\left(\hat{\Pi}_{yx}^{u_y}+\hat{\Pi}_{xy}^{u_y}\right)$$

$$\Pi_{yy} = \frac{p}{N_\delta}\left(\hat{\Pi}_{yy}^{v_y}+\hat{\Pi}_{yy}^{u_y}+\hat{\Pi}_{yy}^{w_y}\right) \quad \Pi_{xz} = \Pi_{zx} = \frac{p}{N_\delta}\left(\hat{\Pi}_{zx}^{u_z}+\hat{\Pi}_{xz}^{w_x}\right)$$

$$\Pi_{zz} = \frac{p}{N_\delta}\left(\hat{\Pi}_{zz}^{w_z}+\hat{\Pi}_{zz}^{u_z}+\hat{\Pi}_{zz}^{v_z}\right) \quad \Pi_{yz} = \Pi_{zy} = \frac{p}{N_\delta}\left(\hat{\Pi}_{zy}^{v_z}+\hat{\Pi}_{yz}^{w_y}\right)$$

$$\Delta = \frac{p}{N_\delta}\left(\hat{\Delta}^{u_x}+\hat{\Delta}^{v_x}+\hat{\Delta}^{w_x}+\hat{\Delta}^{v_y}+\hat{\Delta}^{u_y}+\hat{\Delta}^{w_y}+\hat{\Delta}^{w_z}+\hat{\Delta}^{u_z}+\hat{\Delta}^{v_z}\right) .\tag{52}$$

$$Q_x = \frac{p\sqrt{T/(2\varepsilon)}}{N_\delta}\hat{Q}_x \quad Q_y = \frac{p\sqrt{T/(2\varepsilon)}}{N_\delta}\hat{Q}_y \quad Q_z = \frac{p\sqrt{T/(2\varepsilon)}}{N_\delta}\hat{Q}_z$$

### 3.4.2 Coupled solver

The former research of NCCR model by the uncoupled solver preliminarily demonstrated its capacity in one-dimensional and two-dimensional cases. Based on Myong's uncoupled computational method, three-dimensional problems are simplified approximately into three one-dimensional non-interfering problems in x, y, z directions and the computation of stress and heat flux components on a surface is achieved by the two uncoupled solvers(40)-(43). Nevertheless, this method overlooks the interactional effect of three directions in real flows and weakens the coupled effect of non-conserved variables $\left(\Pi_{xx}, \Pi_{xy}, \Pi_{xz}, \Delta, Q_x\right)$ in constitutive relations. The most unsatisfied feature is the computational instability caused by the existence of a phenomenon that density might come to be negative particularly in some expansion flow regions of wake stream.

Since the first uncoupled solver describing the compressional and expansional processes in flows, it seems that some omitted terms in this solver contain significant physical characteristics in a deviation from the local equilibrium condition. Therefore, our work would rather focus on the direct solutions of the nonlinear coupled constitutive equations (27)-(29) by a coupled solver than follow the former route to solve NCCR model's algebraic system (40)-(43) by the uncoupled solver. The dimensionless evolution equations (27)-(29) for a diatomic gas are transformed into the following forms as

$$\hat{\mathbf{\Pi}}:\hat{\mathbf{\Pi}}q\left(c\hat{R}\right)=(1+\hat{\Delta})\hat{\mathbf{\Pi}}:\hat{\mathbf{\Pi}}_0+\hat{\mathbf{\Pi}}:\left[\hat{\mathbf{\Pi}}\cdot\nabla\hat{\mathbf{u}}\right]^{(2)},\tag{53}$$

$$\frac{2\gamma'}{f_b}\hat{\Delta}^2q\left(c\hat{R}\right)=\frac{2\gamma'}{f_b}\hat{\Delta}\hat{\Delta}_0+3\gamma'\hat{\Delta}(\hat{\Delta}\mathbf{I}+\hat{\mathbf{\Pi}}):\nabla\hat{\mathbf{u}},\tag{54}$$

$$\hat{\mathbf{Q}}\cdot\hat{\mathbf{Q}}q\left(c\hat{R}\right)=(1+\hat{\Delta})\hat{\mathbf{Q}}\cdot\hat{\mathbf{Q}}_0+\hat{\mathbf{\Pi}}:\hat{\mathbf{Q}}_0\hat{\mathbf{Q}}.\tag{55}$$

Owning to the computation complexity of the coupled constitutive equations, above equations have to be merged into one formulation firstly by using the Rayleigh-Onsager dissipation function. From tensorial expressions above, we can get

$$\hat{R}^2q\left(c\hat{R}\right)=F,\tag{56}$$

where

$$F=(1+\hat{\Delta})\hat{\mathbf{\Pi}}:\hat{\mathbf{\Pi}}_0+\hat{\mathbf{\Pi}}:\left[\hat{\mathbf{\Pi}}\cdot\nabla\hat{\mathbf{u}}\right]^{(2)}+\frac{2\gamma'}{f_b}\hat{\Delta}\hat{\Delta}_0+3\gamma'\hat{\Delta}(\hat{\Delta}\mathbf{I}+\hat{\mathbf{\Pi}}):\nabla\hat{\mathbf{u}}+(1+\hat{\Delta})\hat{\mathbf{Q}}\cdot\hat{\mathbf{Q}}_0+\hat{\mathbf{\Pi}}:\hat{\mathbf{Q}}_0\hat{\mathbf{Q}}.$$

Here a simple iterative method will be adopted to solve equation(56) as

$$\hat{R}_n=\left[\hat{\mathbf{\Pi}}_n:\hat{\mathbf{\Pi}}_n+\frac{2\gamma'}{f_b}\hat{\Delta}_n^2+\hat{\mathbf{Q}}_n\cdot\hat{\mathbf{Q}}_n\right]^{1/2}$$

$$\hat{R}_{n+1}=\frac{1}{c}\sinh^{-1}\left(\frac{cF_n}{\hat{R}_n}\right)$$

$$\hat{\mathbf{\Pi}}_{n+1}=\left((1+\hat{\Delta}_n)\hat{\mathbf{\Pi}}_0+\left[\hat{\mathbf{\Pi}}_n\cdot\nabla\hat{\mathbf{u}}\right]^{(2)}\right)\frac{\hat{R}_n\hat{R}_{n+1}}{F_n}\ .\tag{57}$$

$$\hat{\Delta}_{n+1}=\left(\hat{\Delta}_0+\frac{3}{2}f_b(\hat{\Delta}_n\mathbf{I}+\hat{\mathbf{\Pi}}_n):\nabla\hat{\mathbf{u}}\right)\frac{\hat{R}_n\hat{R}_{n+1}}{F_n}$$

$$\hat{\mathbf{Q}}_{n+1}=\left((1+\hat{\Delta}_n)\hat{\mathbf{Q}}_0+\hat{\mathbf{\Pi}}_n\cdot\hat{\mathbf{Q}}_0\right)\frac{\hat{R}_n\hat{R}_{n+1}}{F_n}$$

And the initial values of shear stress, excess normal stress and heat flux in above equation are calculated below by the linear values $\mathbf{\Pi}_0$, $\Delta_0$ and $\mathbf{Q}_0$ from NSF model as

$$\hat{R}_0 = \left[ \hat{\mathbf{\Pi}}_0 : \hat{\mathbf{\Pi}}_0 + \frac{2\gamma'}{f_b} \hat{\Delta}_0^2 + \hat{\mathbf{Q}}_0 \cdot \hat{\mathbf{Q}}_0 \right]^{1/2}$$

$$\hat{\mathbf{\Pi}}_1 = \frac{\sinh^{-1}\left(c\hat{R}_0\right)}{c\hat{R}_0} \hat{\mathbf{\Pi}}_0$$

$$\hat{\Delta}_1 = \frac{\sinh^{-1}\left(c\hat{R}_0\right)}{c\hat{R}_0} \hat{\Delta}_0 \qquad (58)$$

$$\hat{\mathbf{Q}}_1 = \frac{\sinh^{-1}\left(c\hat{R}_0\right)}{c\hat{R}_0} \hat{\mathbf{Q}}_0$$

The complete solutions of NCCR model are also considered to be converged when $\left| \hat{R}_{n+1} - \hat{R}_n \right| \le 10^{-5}$.

## 4 Computational results: verification, validation and discussion

Several classical cases are selected to test the capability of NCCR model and its coupled iterative solver, including hypersonic rarefied flows over an infinitely long cylinder in z-direction on two sets of Knudsen numbers covering slip and transitional regimes at 5.48 and 4 Mach numbers respectively as well as a slip complex flow over a hollow cylinder-flare configuration. Before these cases are discussed in detail, a continuum breakdown parameter, the maximum gradient-length-local Knudsen number, which is proposed by Boyd, et al.[32], need to be introduced as

$$Kn_{GLL} = \frac{\lambda}{Q} \left| \frac{dQ}{dl} \right|, \qquad (59)$$

where the derivative is taken in the direction of maximum gradient and Q is a variable of interest such as density, temperature or pressure. Boyd, et al. suggested a value of Kn above 0.05 as a continuum breakdown point[33].

### 4.1 Hypersonic flows around an infinitely long cylinder in z-direction

#### 4.1.1 A slip flow around a 2D cylinder for a monatomic gas

In this subsection, 3D FVM schemes with the coupled solver for NCCR model is validated for a monatomic gas flow around an infinitely long cylinder (radii equals to 1.9mm) in slip regime. The monatomic gas is assumed argon with $s = 0.75$ in the inverse power laws and $c = 1.0179$. The input parameters for the free-stream conditions from the literature[28] are given as

$$\begin{array}{ll} Ma_\infty = 5.48 & Kn_\infty = 0.05 \\ T_\infty = 26.6K & T_w = 293.15K \\ p_\infty = 5Pa & R = 208.16 m^2/\left(\sec^2 \cdot K\right). \\ \Pr = \dfrac{2}{3} & \gamma = \dfrac{5}{3} \\ \eta_{ref} = 2.27 \times 10^{-5}\ \text{N} \cdot s/m^2 & T_{ref} = 300K \end{array} \qquad (60)$$

Notice that the computational model (23)-(25) can reduce to the model for monatomic gases when the evolution equation (24) for excess normal stress is left out and the bulk viscosity vanishes. We take this case into consideration in order to compare the results computed by the coupled solver for NCCR model with the canonical NCCR results from Myong's uncoupled solver. The capability of NCCR model for monatomic gases in the flow regimes far away from equilibrium is also validated. Typical structured meshes in the cylinder's symmetrical plane are depicted in Figure 1 with 120 and 60 points placed in circumferential and radial directions of the cylinder respectively. Maxwell-Smoluchowski slip and jump boundary conditions(39) are applied at the solid surface.

Figure 2 shows contour of the maximum gradient-length-local Knudsen number computed by NCCR model. It can be seen that continuum breakdown occurs obviously inside the stand-off shock, near the solid surface and in the wake region of the cylinder. A stagnation region near the frontal part of the cylinder is in detailed analysis. The normalized densities and temperatures computed by NSF, NCCR, Boltzmann and DSMC methods along the stagnation streamline are compared in Figure 3 and Figure 4. The DSMC results come from the sophisticated DSMC code[34] with full tangential momentum and thermal accommodation coefficients for slip and jump boundary conditions. The Boltzmann results in Figure 3 are gas kinetic scheme results of nonlinear model Boltzmann equations by Yang and Huang[35]. It is worth mentioning that there is a significant discrepancy in the normalized density and temperature profiles along the stagnation line of the cylinder between the near-local-equilibrium NSF results and the far-from-equilibrium NCCR/DSMC/Boltzmann results. The NCCR model yields good results in better agreement with DSMC/Boltzmann data than the conventional NSF model. At the same time it can be also noticed that non-negligible discrepancies occur in the normalized density and temperature profiles computed by two different solvers for the solution of NCCR model. NCCR by the coupled solver yields better results in the prediction of the location and inner profiles of the bow shock than the uncoupled solver. Especially in Figure 3, the FVM-NCCR density profile predicted by the coupled solver yields almost perfect agreement with the Boltzmann results. Numerical results computed by high-order discontinuous Galerkin method under the limit of the uncoupled computational method do not distinguish its outstanding capability from the second-order finite volume method. It means that the significant flow characteristics left out by the uncoupled solvers cannot be supplemented by high-order numerical schemes accurately. However, an appropriate coupled convergent solver has a significant effect on the stability and accuracy for the solution of NCCR model.

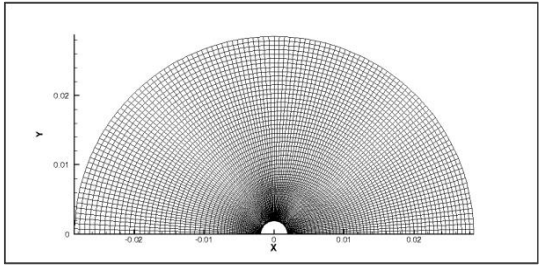

Figure 1 Structured meshes in the cylinder's symmetrical plane

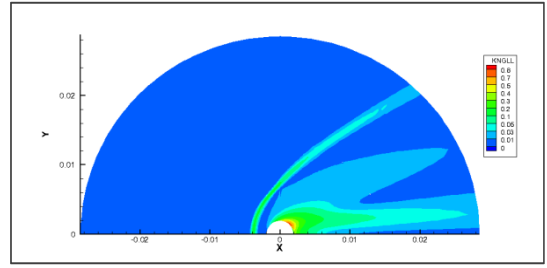

Figure 2 Gradient length local Knudsen contour

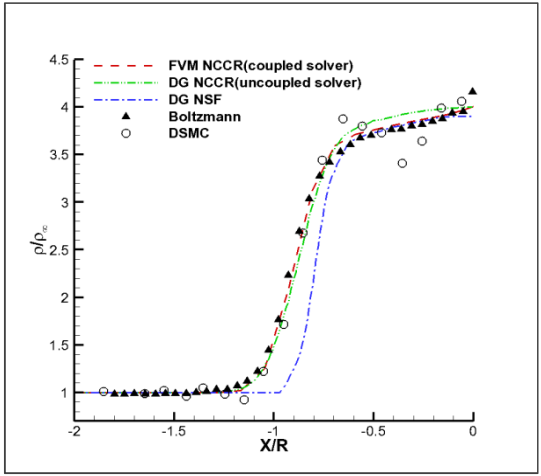

Figure 3 Normalized density distribution along the normalized stagnation line

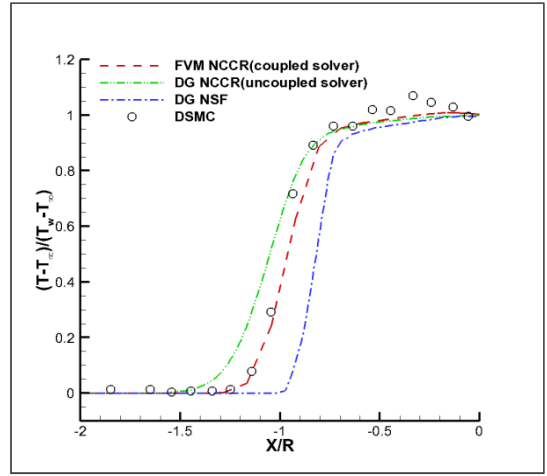

Figure 4 Normalized temperature distribution along the normalized stagnation line

### 4.1.2 A transitional flow over a 2D cylinder for a diatomic gas

A transitional flow over a 2D cylinder for a diatomic gas is also calculated by NCCR model with new developed coupled solver. The radius of the cylinder is 0.05m. The diatomic gas is assumed nitrogen with $s = 0.74$ in the inverse power laws and $c = 1.02029$. The free flow conditions are given as

$$
\begin{aligned}
&Ma_\infty = 4 &&Kn_\infty = 0.16 \\
&U_\infty = 1412.5 m/s &&n_\infty = 1.0 \times 10^{20}/m^3 \\
&T_\infty = 300K &&T_w = 500K \\
&\mathrm{m}_{N_2} = 4.65 \times 10^{-26} \mathrm{kg} &&R = 296.7 m^2/\left(\sec^2 \cdot K\right) \\
&\mathrm{Pr} = 0.72 &&\gamma = 1.4 \\
&\eta_{ref} = 1.656 \times 10^{-5} \mathrm{N} \cdot s/m^2 &&T_{ref} = 273K
\end{aligned}
\tag{61}
$$

The detailed contour comparisons of Mach number, pressure, total average temperature and continuum breakdown parameter between NSF and NCCR are made respectively in Figure 5. As shown in these contours, the typical flow structures include a compressive stand-off shock, a stagnation region near the frontal part of the cylinder and a gaseous expansion region in the wake of the cylinder. It can be seen from the contours of $Kn_{GLL}$ that these flow regions are also the domains far removed from the thermodynamic equilibrium. The $Kn_{GLL}$ number exceeds the critical value of 0.05 in large part of the computational domain for both NSF and NCCR. It is worthwhile pointing out that the breakdown parameter for NCCR even goes up to 10 in the wake near the solid surface which can be considered as free molecular regime. In these continuum-breakdown regions, NSF equation will lose its accuracy and be no longer validated. The contours also illustrate that the shock predicted by NCCR is thicker than that by NSF and the shock standoff distance is removed farther away from the stagnation point by NCCR than by NSF. The DSMC data of the distribution of Mach number, pressure, total average temperature and density along the stagnation line is provided by[36]. One feature highlighted from Figure 6-Figure 9 is that NCCR model yields better agreement with DSMC data in the prediction of the shock standoff distance than NSF. However, there are some non-negligible discrepancies in the inner profiles of the bow shock between DSMC and NCCR results. It implies that the present M-S slip boundary condition may not be well suitable for NCCR model or that NCCR model as a simplified theory of Eu's generalized hydrodynamic equations still overlooks some key features. Therefore, more effort will be paid on these problems in future work.

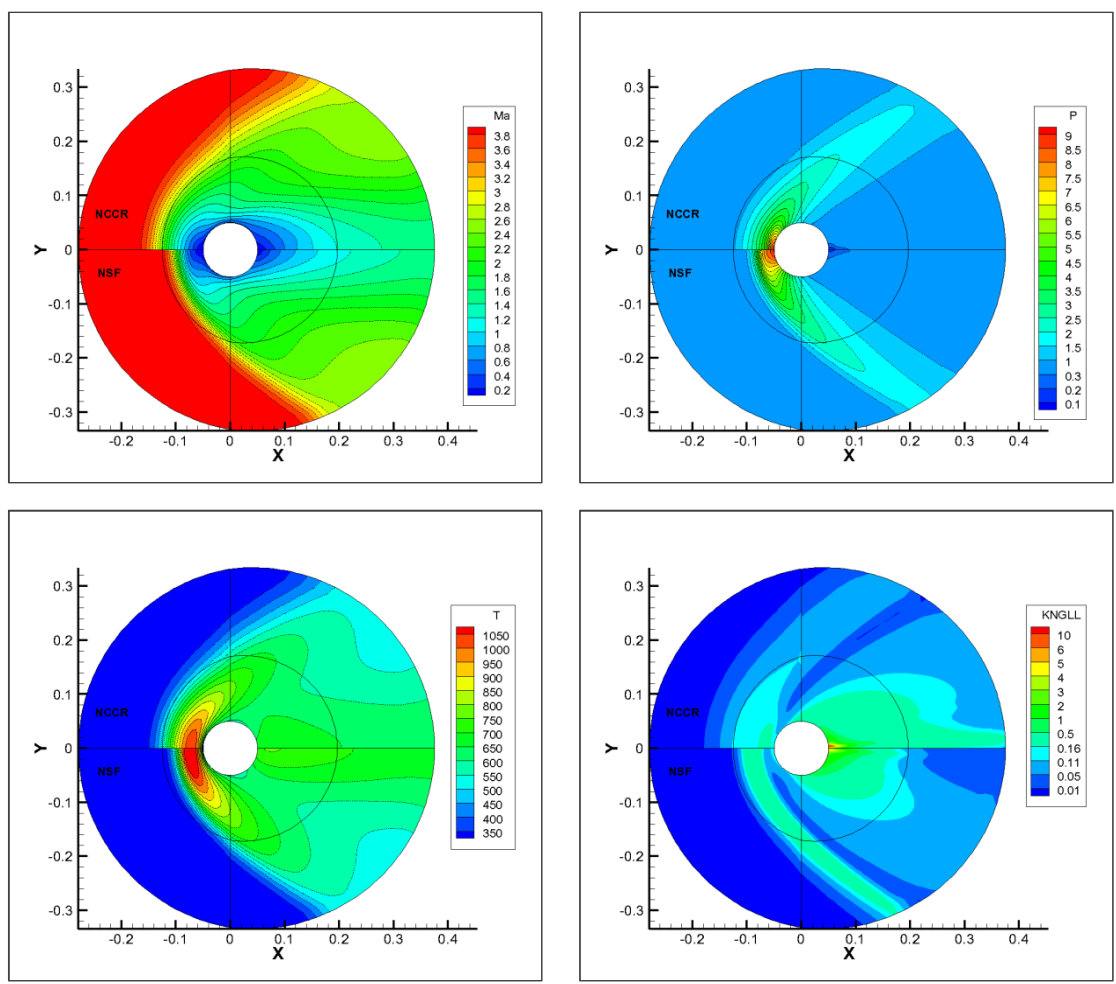

Figure 5 The detailed contour comparisons of Mach number, pressure, total average temperature and continuum breakdown parameter between NSF and NCCR

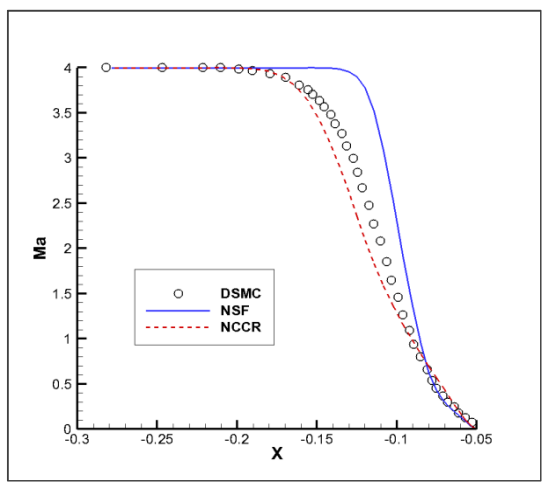

Figure 6 Mach number distribution along the stagnation line

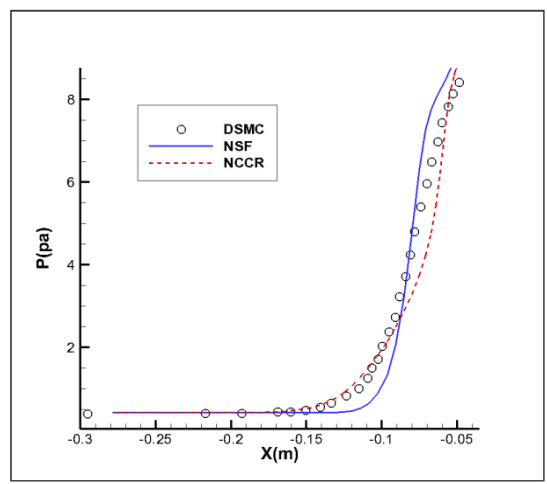

Figure 7 Pressure distribution along the stagnation line

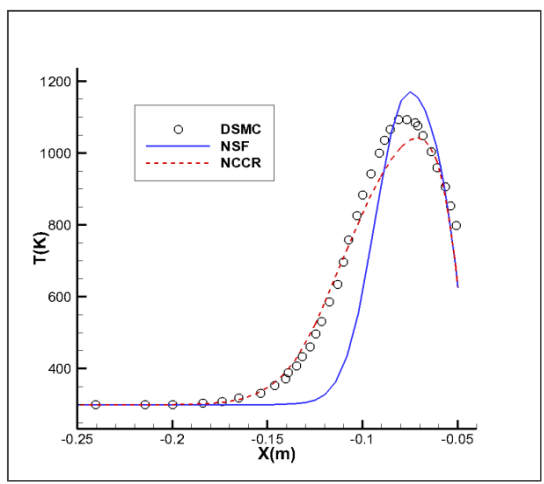

Figure 8 Temperature distribution along the stagnation line

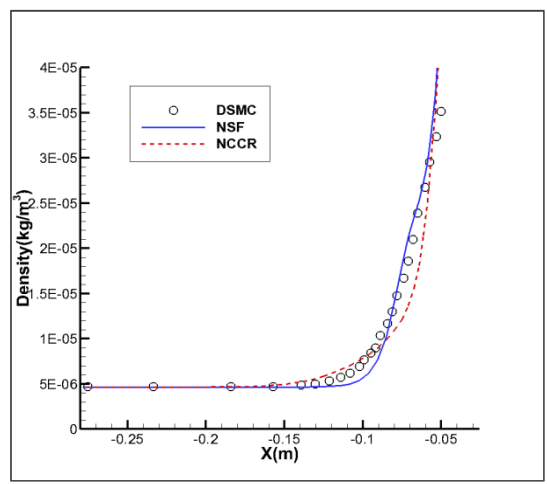

Figure 9 Density distribution along the stagnation line

## 4.2 A complex flow past the hollow cylinder-flare configuration

Before being employed to predict the complex flow of hypersonic vehicles in near-space regime which may be difficult for ground-based facilities, new computational models need to successfully meet some verification tests firstly. In the last testing case, the data of winds tunnel experiments made in the CUBRC LENS facility[37] is available to validate the computational model and this case has also been extensively studied in open literatures [38-41] containing a large amount of DSMC and NS validation data.

The configuration of a hollow cylinder in conjunction with a $30^{\circ}$ conical flare is depicted in Figure 10. It is worthwhile mentioning that the sharp leading edge separates the free stream and only the external flow of this configuration is taken into account here because the internal flow does not interact with the external flow. The working diatomic gas is assumed pure nitrogen with $\Pr = 0.72$, $\gamma = 1.4$, $c = 1.02029$ and $s = 0.74$. The specific inputs for the free-stream conditions are listed as

$$L = 0.1017m \qquad \text{U}_\infty = 2301.7 m/s$$
$$T_\infty = 118.2K \qquad \rho_\infty = 9.023 \times 10^{-4} kg/m^3$$
$$T_w = 295.6K \qquad R = 296 m^2/\left(\sec^2 \cdot K\right) \qquad . \tag{62}$$
$$Kn_\infty = 6.5 \times 10^{-4} \qquad Ma_\infty = 10.4$$
$$\eta_{ref} = 1.656 \times 10^{-5} \text{ N} \cdot s/m^2 \quad T_{ref} = 273K$$

Notice that the free-stream condition is non-equilibrium in CUBRC Run 14 as the translational, rotational and vibrational temperatures are at different time scales. However, since the NCCR model at present does not include these physical non-equilibrium effects, no transitional, rotational and vibrational energy exchange for a diatomic gas is considered in our simulation. According to Myong[24], the excess normal stress associated with the bulk viscosity of a diatomic gas could be introduced to describe simply the rotational non-equilibrium effect in some flow regimes where the rotational relaxation is faster than the hydrodynamic scale. Our present work here mainly focus on that in what range of flow regime NCCR model without these additional physical effect models can be capable of calculating. It is worth mentioning again that the following NCCR results are computed by the new coupled solver.

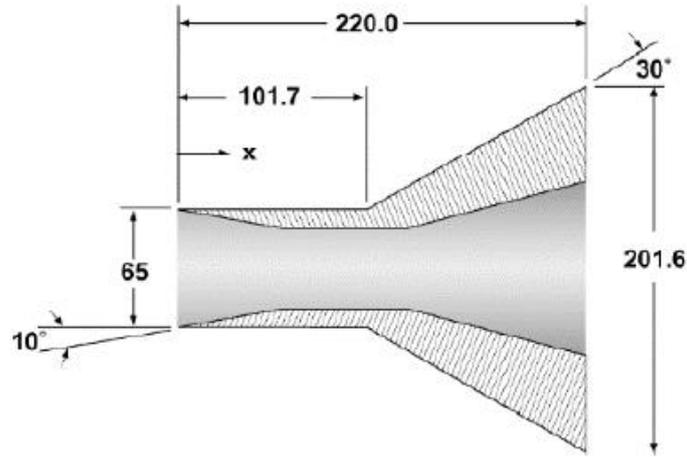

Figure 10 Schematic of hollow cylinder-flare configuration (units in millimeters)[42]

Before we start to study the flow pass though the configuration, gradient-length-local Knudsen contour is shown firstly. As it is displayed in Figure 11, continuum breakdown occurs inside oblique shock wave above the flare and the region between sharp leading edge and the separation region. The size of the separation and re-attachment region can be observed clearly in Figure 12. The Mach number contour which is calculated by NCCR model demonstrates its capability of simulating these hypersonic flows fairly well.

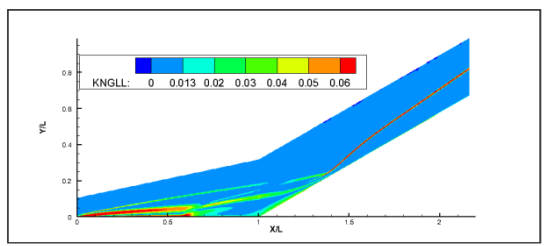

Figure 11 Gradient length local Knudsen contour

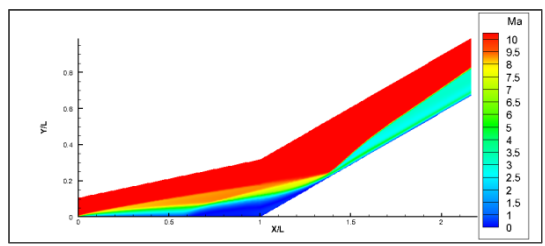

Figure 12 Mach number contour

Since the separation point is at about x/L=0.5, flow fields along the lines normal to the cylinder body at x/L=0.5 and 1 are studied. Figure 13 shows the detailed comparisons for density and velocity properties of DSMC, NSF and NCCR along these two lines. Notice that there is no comparison for temperature here now that the present NCCR model has not taken rotational and vibrational non-equilibrium effect into account. The DSMC and NSF results both come from the literature[39] and the DSMC results were computed by a parallel optimized code named MONACO. As displayed on the left of Figure 13, the comparisons show that the NCCR results computed by the coupled iterative solver are in better agreement with the DSMC results than NSF results along the line normal to the cylinder at x/L=0.5 which is considered to be removed from local thermodynamic equilibrium based on continuum breakdown parameter $Kn_{GLL}$. In the right profiles of Figure 13, NCCR results are only in qualitative agreement with DSMC results, but excellently capture a small discontinuity predicted by DSMC around the shock region which is not captured by NSF. Boyd, et al.[39] pointed out that DSMC did provide correct solution before the separation point, but the accuracy at the conjunction of the cylinder and flare(x/L=1) was questionable. The obvious deviation between NCCR and DSMC results makes it difficult to judge the performance of former one and its unknown cause remains worthwhile to be put on future's research. In Figure 14, numerical results are compared with the experimental data for non-dimensionalized heat flux coefficient along the body surface. The heat flux coefficient is defined as $C_q = q_w / 0.5 \rho_\infty U_\infty^3$. It is very encouraging that the NCCR solution is in outstanding agreement with the experimental data especially in the separation and re-attachment points (about x/L=0.6 and 1.4). In Figure 14, good performance of NCCR model in predicting the size of the separation and re-attachment region which is over-predicted by DSMC and under-predicted by NSF is also demonstrated. From the analysis above, the NCCR model is capable of simulating characteristic hypersonic flows fairly well.

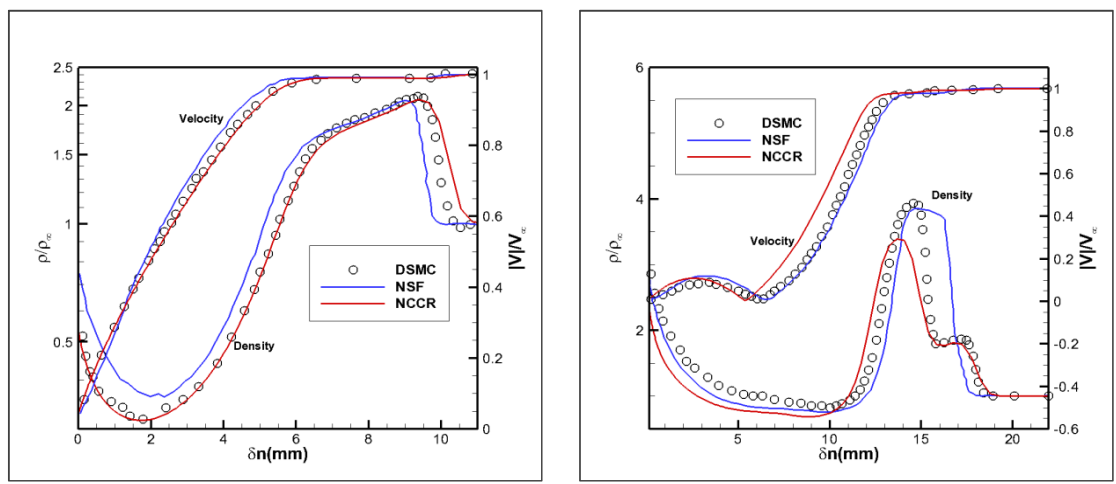

Figure 13 density and velocity profiles along the line normal to the cylinder at x/L=0.5(left) and x/L=1(right)

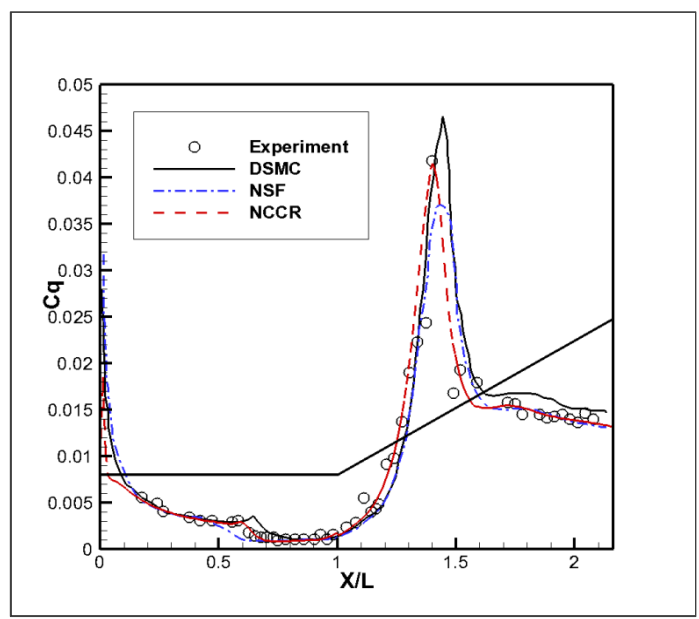

Figure 14 comparison of surface heat flux non-dimensionalized coefficient

## 5 Conclusions

The main emphasis of this paper is placed on the development of a reliable coupled computational method for nonlinear coupled constitutive relations (NCCR) in conjunction with three-dimensional FVM schemes and the further extension of NCCR model into three-dimensional computations. As a step toward developing the coupled solver, the NCCR model, derived from Eu's generalized hydrodynamic equations has been numerically investigated.

Numerical simulations of a slip gas flow around a 2D cylinder for a monatomic gas are conducted firstly. It is concluded that the coupled computational method with three-dimensional upwind FVM schemes for the solution of NCCR model in present work is capable of simulating these far-from-equilibrium flows fairly well. The detailed comparisons show that the flow properties (density and temperature) along the stagnation line computed by the coupled solver for NCCR model match better with DSMC/Boltzmann data than that by the uncoupled solver.

Subsequently, based on the new coupled solution for NCCR model, a transitional flow over another 2D cylinder and a slip flow past a 3D hollow cylinder-flare configuration for a diatomic gas are on detailed researches. The NCCR model is shown to yield some quantitative agreement with experimental and DSMC data in the prediction of hypersonic and rarefied flows compared with linear constitutive relation NSF. The successful applications in these far-from-equilibrium cases have implied the potential of NCCR model as an alternative to more computationally intensive DSMC and less accurate NSF linear constitutive relations in prediction of hypersonic and rarefied flow.

However, there are also some non-negligible unsatisfactory disagreement between NCCR and DSMC data in some flow regions, such as the inner profiles of the bow shock for the cylinder and the separation region at the conjunction of the cylinder and flare. These disappointing performances of NCCR model relative to NSF results imply three crucial problems of whether the present slip boundary condition is proper for this new high-order computational model, whether the rotational and vibrational equilibrium assumption without any additional high-temperature gas effect models is suitable, and to what extent the NCCR model as a simplified theory of Eu's generalized hydrodynamic equations is valid. With the aim of answering the aforementioned questions, more effort will be paid in our future research.

## Acknowledgments

This research was fund by the National Basic Research Program of China (Grant NO. 2014CB340201), the National Natural Science Foundation of China (Grant NO.11502232) and the Fundamental Research Funds for the Central Universities. Meanwhile, the authors would also like to acknowledge the guidance from Prof. Xiao at Northwestern Polytechnical University in China and the discussion from Prof. Myong at Gyeongsang National University.